\documentstyle[12pt]{article}
\begin{document}
\title{The effect of temperature on the viscoelastic response
of rubbery polymers at finite strains}

\author{Aleksey D. Drozdov$^{1}$ and Al Dorfmann$^{2}$\\
$^{1}$ Institute for Industrial Mathematics, 4 Hanachtom Street\\
84311 Beersheba, Israel\\
$^{2}$ Institute of Structural Engineering, 82 Peter Jordan Street\\
1190 Vienna, Austria}
\date{}
\maketitle

\begin{abstract}
Constitutive equations are derived for the viscoelastic response
of rubbery polymers at finite strains.
A polymer is thought of as a network of long chains connected to
temporary junctions.
At a random time, a chain detaches from a junction, which is treated
at transition from its active state to the dangling state.
A dangling chain at a random time captures a new junction in
the vicinity of its free end and returns to its active state.
Breakage and reformation of long chains are modeled as thermo-mechanically
activated processes.
Stress--strain relations for a rubbery polymer are developed using the
laws of thermodynamics.
Adjustable parameters in the model are found by fitting observations
in uniaxial tensile tests for a carbon black filled
rubber at various temperatures.
Fair agreement is demonstrated between experimental data and results of
numerical simulation.
\end{abstract}

\section{Introduction}

This paper is concerned with the nonlinear viscoelastic
behavior of rubbery polymers at finite strains.
The viscoelastic and viscoplastic response of rubbery polymers
has been the focus of attention in the past decade \cite{GS91}--\cite{TBP00}.
This may be explained by the necessity to develop constitutive equations
for rubber-like materials which may be employed for the numerical
solution of applied engineering problems, on the one hand,
and by a number of peculiarities in the behavior of rubber
which cannot be adequately predicted by conventional
stress--strain relations, on the other.
The time-dependent response of industrial rubbery polymers in conventional
mechanical tests is affected by a number of accompanying phenomena:
\begin{enumerate}
\item
Plastic deformation and micro-damage (initiation and development
of shear bands and crazes) \cite{GS91,RW97,KB90}.

\item
Meso-scale damage (cavitation) and void formation \cite{SG99,DB00}.

\item
Mechanically induced crystallization of semicrystalline polymers
\cite{ZB97,LB99,HMS00}.

\item
Slip \cite{GS97} and fragmentation \cite{HHL98} of crystallites driven by
network stretching.

\item
Mullins's effect \cite{GS92,JB93a,JB93b}.

\item
Payne's effect \cite{HVH96}.

\item
Strain induced evolution of the internal structure \cite{LAP93}
and orientational hardening \cite{ZB97,LB99,WG93}.

\item
Reinforcement of rubbery polymers by carbon black and silica
\cite{SC91,MTD93,Ulm95,HVH96,KR98,TBP00}
and fractal aggregation of particles \cite{WRC93}.
\end{enumerate}
Although this list is far from being exhausted, even these items
show that the existing micro-mechanical models are oversimplified
to quantitatively predict the time-dependent response of rubbers.
Even if we disregard these phenomena for an instant,
a number of unresolved problems is found out which are associated
with the description of rubbery polymers at the micro-level.

This study focuses on the following questions:
\begin{enumerate}
\item
According to the entropic theory of rubber \cite{Tre75,Fer80},
a polymer may be treated as a network of long chains connected
to junctions.
A junction is thought of as a crosslink between two (or more) monomers
belonging to different chains.
A chain between two junctions consists of a large number of strands
(statistically independent elements of a macromolecule)
which are modeled as rigid rods.
This implies that the mechanical energy of a chain vanishes
and the main contribution to the free energy of a chain is
provided by the configurational entropy (that characterizes
the number of configurations available for a chain which is treated
as a random walk on a three-dimensional lattice) \cite{Fer80}.
This approach correctly predicts the response of polymeric gels
near the sol--gel transition point.
Its application to more dense media (e.g., to polymeric melts)
results in large deviations between experimental data and results
of numerical analysis.
To avoid these discrepancies, a mechanical energy of chains is
introduced ``by hand'' into the formula for the free energy \cite{DE86}
(which is tantamount to treatment of strands as linear or nonlinear
elastic springs).
On the other hand, a conventional standpoint is that the concept
of entropic elasticity is inapplicable to glassy polymers,
whose response is associated with cooperative relaxation in
micro-domains (regions of relatively high density comprising
dozens of strands which belong to different chains and which
rearrange simultaneously because of large-angle reorientation
of strands) \cite{AG65,Dyr95,MB96,Sol98}.
An open question is about the correspondence between the entropic
contribution into the free energy of a rubbery polymer and
the contribution of its mechanical energy at various temperatures $T$
(in particular, near the glass transition temperature $T_{\rm g}$).

\item
Changes in instantaneous elastic moduli of rubbery polymers
with temperature are conventionally described by two mechanisms
at the micro-level.
According to the first \cite{Fer80}, an increase in
temperature results in thermal expansion of polymers, which is
equivalent to an increase in the average distance between sequential
monomers in a chain and between neighboring chains.
The growth of this distance implies a decrease in the intramolecular
and intermolecular forces which is observed as a reduction
of instantaneous elastic moduli at the macro-level.
According to the other approach \cite{Str90}, the rubbery state of
a polymer is characterized by a large number of entanglements between
long chains.
At elevated temperatures, entanglements do not impose
restrictions on micro-motion of individual strands (local constrains),
but serve only as global topological constrains for the motion
of macromolecules.
With a decrease in temperature (or an increase in the strain rate),
the number of entanglements grows that restrict
the micro-motion of particular strands of long chains,
which is equivalent to transformation of these entanglements
into temporary crosslinks.
An increase in the number of temporary crosslinks is tantamount
to a decrease in the average number of strands in a chain.
The question is which of these two mechanisms is more adequate
for the description of observations in mechanical tests.

\item
According to conventional theories of rubber elasticity \cite{Tre75},
a polymer is treated as an ensemble of flexible chains
(the energy associated with their bending is neglected).
The account for mechanical energy of elongation of chains
in the formula for the free energy
implies the question whether a model of semiflexible chains
(with finite bending rigidities) is more adequate for
the description of rubbers, and in what way the bending rigidity
should be introduced into the constitutive equations.
\end{enumerate}

This paper is an attempt to shed some light on these issues.
To describe the time-dependent behavior of rubbery polymers,
we employ the concept of transient networks.
This theory was proposed in the seminal work \cite{GT46}
and developed in \cite{Yam56,Lod68,TE92}.
In the past decade, the theory of temporary networks was
widely used to describe various observations
from shear thickening of polymer solutions \cite{Wan92}
to nonlinear relaxation of telechelic polymers \cite{SJB00}.
The original theory was refined to account for
(i) the effect of mechanical factors on the rates of breakage
and reformation \cite{PTT77,FL81,PB88,AO95}
and (ii) the finite rate of stress relaxation in dangling
chains \cite{Pal97}.
All previous studies, however, treated transient networks as consisting
of chains with an identical number of strands.
This hypothesis substantially simplifies the analysis, but it
imposes unreasonable limitations on the model.
Molecular dynamics simulation \cite{BS99} reveals that the
chain length dramatically affects its relaxation rate.
Based on this observations, we assume a transient network
to be composed of long chains with different lengths
and characterize the network by the probability density
of chains with various numbers of strands.
This refinement provides a way to distinguish changes in the
viscoelastic response driven by (i) an increase in the elastic modulus
of a strand and (ii) the growth of the number of temporary
crosslinks.

The exposition is organized as follows.
Section 2 deals with kinetic equations for the rates of breakage
and reformation.
The strain energy density of a transient network is determined
in Section 3.
Section 4 is concerned with constitutive equations for
a temporary network of long chains.
The stress--strain relations are applied to describe uniaxial extension
of a specimen in Section 5.
Several hypotheses regarding (i) the distribution of chains with various
number of strands and (ii) the effect of the number of strands
on the rate of a chain's breakage are introduced in Section 6.
The stress--strain relations are verified in Section 7 by comparing
results of numerical simulation with experimental data for carbon
black filled rubber in static and dynamic tests.
Some concluding remarks are formulated in Section 8.

\section{The concept of temporary networks}

A rubbery polymer is modeled as a network of long chains
(with various numbers of strands) bridged by temporary junctions.
A chain whose ends are connected to separate junctions
is treated as an active one.
Snapping of an end of a chain from a junction is thought of as
its breakage (transition from its active state to the dangling state).
When a dangling chain captures a junction, a new active chain
arises \cite{TE92}.
Breakage and reformation of active chains are treated as thermally
activated processes: attachment and detachment events
occur at random times as they are driven by thermal fluctuations.

A chain is determined by two parameters: the number of strands,
$n=1,2\ldots$, and the end-to-end vector, $\bar{R}$.
The distribution of chains with various numbers of strands is determined
by the probability density $p(n)$, which is assumed to be
independent of temperature, $T$, and the strain intensity.
This hypothesis is equivalent to the assumption that chains do not fall
to pieces under loading.

The kinetics of reformation is described by the function
$X(t,\tau,n)$, which equals the number of active
(merged to the network) chains with $n$ strands (per unit mass)
at time $t$ that has last been linked to the network at
instant $\tau\in [0,t]$.
This function entirely determines the current state
of a rubbery polymer at the micro-level.
For example, the quantity $X(t,t,n)$ equals the
number of active chains (per unit mass) with $n$ strands at time $t$.
In particular, $X(0,0,n)$ is the initial concentration
of active chains with $n$ strands,
\begin{equation}
X(0,0,n)=\Xi p(n),
\end{equation}
where $\Xi$ is the total number of active chains per unit mass
(we suppose that $\Xi$ does not change with strains and coincides
with the number of active chains in a stress-free medium).
The amount
\[ \frac{\partial X}{\partial \tau}
(t,\tau,n) \biggl |_{t=\tau} \; d\tau \]
is the number of active chains with $n$ strands (per unit mass)
that have been linked to the network within the interval
$[\tau,\tau+d\tau ]$;
the quantity
\[ \frac{\partial X}{\partial \tau} (t,\tau,n)\;d\tau \]
determines the number of these chains that did not
break during the interval $[\tau, t]$;
the amount
\[ -\frac{\partial X}{\partial t} (t,0,n)\;dt \]
is the number of active chains (per unit mass) that detach from the
network (for the first time) within the interval $[t,t+dt ]$,
and the quantity
\[ -\frac{\partial^{2} X}{\partial t\partial \tau} (t,\tau,n)\;dtd\tau \]
is the number of long chains (per unit mass) that has last been linked
to the network within the interval $[\tau,\tau+d\tau ]$
and leave the network (for the first time after merging)
during the interval $[t,t+dt ]$.

The kinetics of evolution for a network is determined
by the relative rate of breakage of active chains, $\Gamma(t,n)$,
and by the rate of merging of dangling links with the
network, $\gamma(t,n)$.
The assumption that $\Gamma$ and $\gamma$ depend on
the number of strands $n$ distinguishes the present model from previous
ones where all chains are treated as composed of an identical number
of strands.
The dependence of the rates of breakage and reformation on time
reflects the effect of the current strain intensity on these
parameters at time-varying loading.

The relative rate of breakage, $\Gamma$, is defined
as the ratio of the number of active chains broken per unit time
to the total number of active chains,
\begin{eqnarray}
\Gamma(t,n) &=& -\frac{1}{X(t,0,n)}
\frac{\partial X}{\partial t}(t,0,n),
\nonumber\\
\Gamma(t,n) &=& -\biggl [ \frac{\partial X}{\partial \tau}
(t,\tau,n) \biggr ]^{-1}
\frac{\partial^{2} X}{\partial t\partial \tau}(t,\tau,n).
\end{eqnarray}
In the general case, $\Gamma$ depends on the instant $\tau$
when a chain has been last bridged to the network (before its detachment)
and on the guiding vector $\bar{l}$ (the unit vector
whose direction coincides with the end-to-end vector of the chain
at the time $\tau$).
We neglect these dependencies, which is equivalent to the assumption
that the network is homogeneous (the rate of breakage
for a chain is independent of the instant of its connection to
the network) and isotropic (the rate of breakage for a chain
is independent of its direction at the instant of merging).

In agreement with conventional theories of temporary networks \cite{TE92},
the rate of merging, $\gamma$, is defined as the number
of dangling chains (per unit mass) bridged to the network per unit time
\begin{equation}
\gamma(t,n)=\frac{\partial X}{\partial \tau}
(t,\tau,n)\biggl |_{\tau=t}.
\end{equation}
The hypothesis about the isotropicity of the network implies that
$\gamma$ is independent of the guiding vector $\bar{l}$.

We consider a model with two states (active and dangling) of a long chain.
A dangling chain is treated as a chain where stresses totally relax
after its detachment (no memory about previous deformations),
whereas the active chain is a chain that preserves entire memory
about the macro-strain at the instant of its connection with the network.
One can hypothesize about an intermediate state between these two
extremities, when the interval between the detachment event and
the subsequent attachment to the network is too small
for total relaxation of stresses,
but is sufficiently large to ensure that a part of these stresses
does relax \cite{Pal97}.
This is equivalent to the assumption that a partial memory preserves
in a chain about the history of its deformation.
For the sake of simplicity, we exclude this case from the consideration.
However, because preserving a partial memory about the loading history
seems quite natural, we distinguish the rate of transition
from the active state to the dangling state, $\Gamma$,
and the attempt rate, $\Gamma_{\ast}$, which is determined as the relative
rate of slippage from sticky junctions for ends of active chains.
It is assumed that $\Gamma_{\ast}$ substantially exceeds the rate of
breakage, $\Gamma$, which means that majority of active chains
detach from the network and, immediately afterwards (before stresses
totally relax), merge with some junctions.

Equations (2) may be treated as ordinary differential equations for
the function $X$.
Integration of Eq. (2) with initial conditions (1) and (3) implies that
\begin{eqnarray}
X(t,0,n) &=& \Xi p(n)\exp
\biggl [ -\int_{0}^{t} \Gamma(s,n) ds\biggr ],
\nonumber\\
\frac{\partial X}{\partial \tau}(t,\tau,n) &=&
\gamma(\tau,n)\exp \biggl [-\int_{\tau}^{t} \Gamma(s,n) ds \biggr ].
\end{eqnarray}
To exclude the rate of reformation $\gamma$ from Eq. (4),
we postulate that the concentrations of active and dangling
chains remain constant and independent of the strain level.
This means that the number of chains broken per unit time
coincides with the number of chains merging with the network
within the same interval.
The number of long chains detached from the network during the interval
$[t,t+dt ]$ equals the sum of the number of initial chains
(i.e., not broken until time $t$) that slip from temporary junctions
\[ -\frac{\partial X}{\partial t}(t,0,n)\; dt \]
and the number of chains connected with the network within the
interval $[\tau, \tau+d\tau ]$ (for the last time before instant $t$)
and broken within the interval $[t,t+dt ]$
\[ -\frac{\partial^{2} X}{\partial t\partial \tau}(t,\tau,n)\;dt d\tau. \]
This number coincides with the number of chains attached to
the network within the interval $[t,t+dt]$,
\[ \gamma(t,n)\; dt, \]
which results in the balance law
\[ \gamma(t,n)=-\frac{\partial X}{\partial t}(t,0,n)
-\int_{0}^{t} \frac{\partial^{2} X}{\partial t\partial \tau}
(t,\tau,n) d\tau . \]
Substitution of Eq. (4) into this equality implies that
\[ \gamma(t,n) = \Gamma (t,n)\biggl \{
\Xi p(n)\exp \biggl [-\int_{0}^{t} \Gamma(s,n) ds \biggr ]
+\int_{0}^{t} \gamma(\tau,n)
\exp \biggl [-\int_{\tau}^{t} \Gamma(s,n) ds \biggr ] d\tau\biggr \}. \]
To solve this equation, we introduce the notation
\begin{equation}
\tilde{\gamma}(t,n)=\gamma(t,n)\exp \biggl [\int_{0}^{t}
\Gamma(s,n) ds\biggr ],
\end{equation}
and find that
\begin{equation}
\tilde{\gamma}(t,n)=\Gamma(t,n)\biggl [ \Xi p(n)
+\int_{0}^{t}\tilde{\gamma}(\tau,n)d\tau \biggr ].
\end{equation}
Setting $t=0$ in Eq. (6), we obtain
\begin{equation}
\tilde{\gamma}(0,n)=\Gamma(0,n)\Xi p(n).
\end{equation}
Differentiation of Eq. (6) with respect to time results in
\[ \frac{\partial \tilde{\gamma}}{\partial t}(t,n)
=\biggl [ \Gamma(t,n)+\frac{1}{\Gamma(t,n)}
\frac{\partial \Gamma}{\partial t}(t,n)\biggr ]
\tilde{\gamma}(t,n). \]
Integration of this equality from zero to $t$ yields
\begin{equation}
\ln \frac{\tilde{\gamma}(t,n)}{\tilde{\gamma}(0,n)}
=\ln \frac{\Gamma(t,n)}{\Gamma(0,n)}
+\int_{0}^{t}\Gamma(s,n) ds.
\end{equation}
It follows from Eqs. (7) and (8) that
\[ \tilde{\gamma}(t,n)=\Xi p(n)\Gamma(t,n)\exp \biggl [\int_{0}^{t}
\Gamma(s,n) ds \biggr ]. \]
Combining this equality with Eq. (5), we arrive at the formula for
the rate of reformation
\begin{equation}
\gamma(t,n)=\Xi p(n)\Gamma(t,n).
\end{equation}

\section{Strain energy density of a temporary network}

An active chain is thought of as a nonlinear elastic spring
whose position is determined by the unit vector $\bar{l}$
directed along the end-to-end vector $\bar{R}$.
We suppose that the stress in a dangling chain totally relaxes
before this chain captures a new junction and
the natural (stress-free) configuration of a chain merging
with the network at time $\tau$ coincides with
the actual configuration of the network at that instant.

We begin with the calculation of the extension ratio (along the chain),
$\lambda$, for a chain that merged with the network
at time $\tau$ and has not been broken within the interval $[\tau, t]$.
At instant $\tau$, the chain has a small length $\delta$
and connects two junctions at points $A$ and $B$.
At time $t\geq \tau $, the junctions occupy points with the radius vectors
\[ \bar{r}_{A}(t)=\bar{r}_{A}(\tau)+\bar{u}(t,\tau,\bar{r}_{A}(\tau)),
\qquad
\bar{r}_{B}(t)=\bar{r}_{B}(\tau)+\bar{u}(t,\tau,\bar{r}_{B}(\tau)), \]
where $\bar{u}(t,\tau,\bar{r})$ is the displacement vector
at point $\bar{r}$ for transition of the network
from the deformed configuration at time $\tau$
to the deformed configuration at time $t$.
The position of the chain is determined by the end-to-end vectors
\begin{eqnarray}
\bar{R}(\tau) &=& \bar{r}_{B}(\tau)-\bar{r}_{A}(\tau)
=\delta\bar{l},
\nonumber\\
\bar{R}(t) &=& \bar{r}_{B}(t)-\bar{r}_{A}(t)
=\delta\bar{l}
+[\bar{u}(t,\tau,\bar{r}_{B}(\tau))-\bar{u}(t,\tau,\bar{r}_{A}(\tau))]
\nonumber\\
&=& \delta\bar{l}
+[\bar{u}(t,\tau,\bar{r}_{A}(\tau)+\delta\bar{l})
-\bar{u}(t,\tau,\bar{r}_{A}(\tau))].
\end{eqnarray}
Neglecting terms of the second order of smallness compared to $\delta$,
we find from Eq. (10) that
\[ \bar{R}(t)=\delta\bar{l}\cdot [\hat{I}
+\bar{\nabla}(\tau)\bar{u}(t,\tau,\bar{r}_{A}(\tau))], \]
where $\bar{\nabla}(t)$ is the gradient operator
in the actual configuration at instant $t$,
$\hat{I}$ is the unit tensor
and dot stands for the inner product.
In terms of the radius vector $\bar{r}$, this equality reads
\begin{equation}
\bar{R}(t)=\delta\bar{l}\cdot \bar{\nabla}(\tau)\bar{r}(t)
=\Bigl [ \bar{\nabla}(\tau)\bar{r}(t)\Bigr ]^{\top} \cdot \delta\bar{l},
\end{equation}
where $\top$ stands for transpose
and the argument $\bar{r}_{A}$ is omitted for simplicity.
The end-to-end length of the chain, $ds$, is given by
\[ ds^{2}=\bar{R}\cdot \bar{R}. \]
Substitution of Eqs. (10) and (11) into this equality results in
\[ ds^{2}(\tau) = \delta^{2},
\qquad
ds^{2}(t) = \delta^{2}\bar{l}\cdot \hat{C}(t,\tau)\cdot \bar{l}, \]
where
\[ \hat{C}(t,\tau)=\bar{\nabla}(\tau)\bar{r}(t)
\cdot \Bigl [ \bar{\nabla}(\tau)\bar{r}(t)\Bigr ]^{\top} \]
is the relative Cauchy deformation tensor for transition from
the deformed configuration of the network at time $\tau$
to the deformed configuration at time $t$.
The extension ratio $\lambda$ is determined as
\begin{equation}
\lambda(t,\tau,\bar{l})=\frac{ds(t)}{ds(\tau)}
=\left [ \bar{l}\cdot\hat{C}(t,\tau)\cdot\bar{l} \right ]^{\frac{1}{2}}.
\end{equation}
With reference to \cite{TE92}, we suppose that the network is
incompressible and neglect the energy of interaction between long chains.
These hypotheses are based on the assumption that the effect of
excluded volume and other multi-chain effects for an individual chain
are screened by surrounded macromolecules.

The mechanical energy of a network is defined as the sum of
the mechanical energies for elongation of long chains
comprising the network.
This assumption is equivalent to the hypothesis that the bending
energy of semi-flexible chains is small compared to its energy
of elongation and may be neglected in the formula for the
total energy of a transient network.
It is worth noting some recent studies \cite{Mor98a,Mor98b}
where the bending energy is included into the formula for the mechanical
energy of an ensemble of chains.

The mechanical energy of a chain with $n$ strands that merged
with the network at instant $\tau$ and has not been broken until
the time $t$ reads
\[ \tilde{W} \Bigl (\lambda(t,\tau,\bar{l}),n \Bigr ), \]
where $\tilde{W}$ is a smooth function.
We postulate that $\tilde{W}(\lambda,n)$ may be factorized as
\begin{equation}
\tilde{W}(\lambda,n)=\mu(n) W_{0}(\lambda),
\end{equation}
where $\mu(n)$ is the rigidity of a chain with $n$ strands
(with the dimension of energy) and $W_{0}(\lambda)$ is a dimensionless
function of the extension ratio which satisfies the condition
\begin{equation}
W_{0} (1)=0.
\end{equation}
Equation (14) means that the mechanical energy vanishes
for a non-deformed chain.

The mechanical energy (per unit mass) of initial chains
which have not been broken within the interval $[0,t]$ is given by
\begin{equation}
\frac{1}{4\pi} \sum_{n=1}^{\infty} \mu(n) X(t,0,n)
\int_{\cal S} W_{0} \Bigl (\lambda(t,0,\bar{l})\Bigr ) dA(\bar{l}),
\end{equation}
where ${\cal S}$ is the unit sphere in the space of vectors $\bar{l}$
and $dA(\bar{l})$ is the surface element on ${\cal S}$.
The mechanical energy (per unit mass) of long chains
that merged with the network within the interval
$[\tau,\tau+d\tau ]$ and have been connected with the network until
the current time $t$ reads
\begin{equation}
\frac{1}{4\pi} \sum_{n=1}^{\infty} \mu(n)
\frac{\partial X}{\partial \tau}(t,\tau,n) d\tau
\int_{\cal S} W_{0} \Bigl (\lambda(t,\tau,\bar{l})\Bigr ) dA(\bar{l}).
\end{equation}
Summing expressions (15) and (16), we arrive at the formula
for the strain energy density (per unit mass)
of the network of long chains
\begin{eqnarray}
U(t) &=& \frac{1}{4\pi} \sum_{n=1}^{\infty}  \mu(n) \biggl [ X(t,0,n)
\int_{\cal S} W_{0} \Bigl (\lambda(t,0,\bar{l})\Bigr ) dA(\bar{l})
\nonumber\\
&&+ \int_{0}^{t} \frac{\partial X}{\partial \tau}(t,\tau,n) d\tau
\int_{\cal S} W_{0} \Bigl (\lambda (t,\tau,\bar{l})\Bigr ) dA(\bar{l})
\biggr ].
\end{eqnarray}
It follows from Eq. (12) that for an isotropic network, the expression
\[ \int_{\cal S} W_{0} \Bigl (\lambda (t,\tau,\bar{l})\Bigr ) dA(\bar{l}) \]
depends on the principal invariants of the relative Cauchy
deformation tensor, $\hat{C}(t,\tau)$, only.
Because the third principal invariant of this tensor equals unity for
an incompressible medium, we obtain
\begin{equation}
\int_{\cal S} W_{0} \Bigl (\lambda (t,\tau,\bar{l})\Bigr ) dA(\bar{l})
=4\pi W\Bigl (I_{k}(t,\tau)\Bigr ),
\end{equation}
where $I_{k}(t,\tau)$ is the $k$th principal invariant of $\hat{C}(t,\tau)$
$(k=1,2)$, and $W=W(I_{1},I_{2})$ is the average (over end-to-end vectors)
mechanical energy (at time $t$) of chains that merge with the network
at instant $\tau$.
Equations (13) and (18) imply that
\begin{equation}
W(3,3)=0,
\end{equation}
where the conditions $I_{1}=3$ and $I_{2}=3$ describe the stress-free
state of a chain.

By analogy with Eq. (18), we write
\begin{equation}
\int_{\cal S} W_{0} \Bigl (\lambda(t,0,\bar{l})\Bigr ) dA(\bar{l})
=4\pi W\Bigl (I_{0k}(t)\Bigr ),
\end{equation}
where $I_{0k}(t)$ is the $k$th principal invariant of the Cauchy deformation
tensor for transition from the initial (stress-free) configuration of
the network to its deformed configuration at time $t$.

Substitution of Eqs. (18) and (20) into Eq. (17) results in the formula
\begin{equation}
U(t) = \sum_{n=1}^{\infty}  \mu(n)
\biggl [ X(t,0,n)W\Bigl (I_{0k}(t)\Bigr )
+ \int_{0}^{t} \frac{\partial X}{\partial \tau}(t,\tau,n)
W\Bigl (I_{k}(t,\tau)\Bigr ) d\tau \biggr ].
\end{equation}

Our objective now is to calculate the derivative of the function $U$
with respect to time.
It follows from Eqs. (2), (19) and (21) that
\begin{equation}
\frac{dU}{dt}(t)=J_{1}(t)-J_{2}(t),
\end{equation}
where
\begin{equation}
J_{1}(t) = \sum_{n=1}^{\infty} \mu(n)\biggl [
X(t,0,n)\frac{dW}{dt}(I_{0k}(t))
+\int_{0}^{t} \frac{\partial X}{\partial \tau}(t,\tau,n)
\frac{\partial W}{\partial t}(I_{k}(t,\tau))d\tau \biggr ],
\end{equation}
and
\begin{equation}
J_{2}(t)= \sum_{n=1}^{\infty}  \mu(n)\Gamma(t,n)
\biggl [ X(t,0,n)W\Bigl (I_{0k}(t)\Bigr )
+ \int_{0}^{t} \frac{\partial X}{\partial \tau}(t,\tau,n)
W\Bigl (I_{k}(t,\tau)\Bigr ) d\tau \biggr ].
\end{equation}
Bearing in mind that the principal invariants of the relative Cauchy
deformation tensor $\hat{C}(t,\tau)$ coincide with the principal
invariants of the Finger tensor for transition from the deformed
configuration at time $\tau$ to the deformed configuration at time $t$,
\[ \hat{F}(t,\tau)=\Bigl [\bar{\nabla}(\tau)\bar{r}(t)\Bigr ]^{\top}
\cdot \bar{\nabla}(\tau)\bar{r}(t), \]
and using the chain rule for differentiation, we find that
\begin{equation}
\frac{\partial W}{\partial t}(I_{k}(t,\tau))
=\sum_{m=1}^{2} \frac{\partial W}{\partial I_{m}}(I_{k}(t,\tau))
\frac{\partial I_{m}}{\partial \hat{F}}(\hat{F}(t,\tau)):
\biggl [ \frac{\partial \hat{F}}{\partial t}(t,\tau)\biggr ]^{\top},
\end{equation}
where the colon stands for convolution of tensors.
It is easy to prove that \cite{Dro96}
\[ \frac{\partial I_{1}}{\partial \hat{F}}(\hat{F})=\hat{I},
\qquad
\frac{\partial I_{2}}{\partial \hat{F}}(\hat{F})
=I_{1}(\hat{F})\hat{I}-\hat{F}^{\top}. \]
Substituting these expressions into Eq. (25) and taking into account the
symmetry of the Finger tensor, we obtain
\begin{equation}
\frac{\partial W}{\partial t}
=\biggl [\Bigl (\frac{\partial W}{\partial I_{1}}
+I_{1}\frac{\partial W}{\partial I_{2}}\Bigr )\hat{I}
-\frac{\partial W}{\partial I_{2}}\hat{F}\biggr ]:
\frac{\partial \hat{F}}{\partial t}.
\end{equation}
The derivative of the Finger tensor with respect to time
is given by \cite{Dro96}
\begin{equation}
\frac{\partial \hat{F}}{\partial t}(t,\tau)
=\Bigl [ \bar{\nabla}(t)\bar{v}(t)\Bigr ]^{\top}\cdot \hat{F}(t,\tau)
+\hat{F}(t,\tau)\cdot\bar{\nabla}(t)\bar{v}(t),
\end{equation}
where
\[ \bar{v}(t)=\frac{d\bar{r}}{dt}(t) \]
is the velocity vector.
It follows from Eqs. (26) and (27) that
\begin{equation}
\frac{\partial W}{\partial t}= 2 \biggl [
\Bigl ( \frac{\partial W}{\partial I_{1}}
+I_{1}\frac{\partial W}{\partial I_{2}}\Bigr )\hat{F}
-\frac{\partial W}{\partial I_{2}}\hat{F}^{2} \biggr ]: \hat{D},
\end{equation}
where
\[ \hat{D}(t)
=\frac{1}{2}\biggl [ \Bigl (\bar{\nabla}(t)\bar{v}(t)\Bigr )^{\top}
+\bar{\nabla}(t)\bar{v}(t)\biggr ] \]
is the rate-of-strain tensor.
Substitution of Eq. (28) into Eqs. (22) and (23) results in
\begin{eqnarray}
\frac{dU}{dt}(t) &=& -J_{2}(t)+ 2 \sum_{n=1}^{\infty} \mu(n)
\biggl \{ X(t,0,n)\Bigl [ \psi_{01}(t)\hat{F}_{0}(t)
+\psi_{02}(t)\hat{F}_{0}^{2}(t)\Bigr ]
\nonumber\\
&& +\int_{0}^{t} \frac{\partial X}{\partial \tau}(t,\tau,n)
\Bigl [ \psi_{1}(t,\tau)\hat{F}(t,\tau)
+\psi_{2}(t,\tau)\hat{F}^{2}(t,\tau)\Bigr ] d\tau\biggr \}: \hat{D}(t),
\end{eqnarray}
where
\begin{equation}
\psi_{1}=\frac{\partial W}{\partial I_{1}}
+I_{1}\frac{\partial W}{\partial I_{2}},
\qquad
\psi_{2}=-\frac{\partial W}{\partial I_{2}},
\end{equation}
and the subscript ``zero'' stands for the quantities that describe
transition from the stress-free configuration of the network
(before loading) to its deformed configuration at the current time $t$.

\section{Constitutive equations for a transient network}

The absolute temperature $T$ is assumed to be close to its reference
value $T_{0}$, which implies that the effect of temperature
on material parameters, as well as thermal expansion of the network
may be neglected.
For an incompressible network, the first law of thermodynamics
reads \cite{CG67}
\begin{eqnarray}
\frac{d\Phi}{dt}=\frac{1}{\rho}\left (\hat{\sigma}_{\rm d}:\hat{D}
-\bar{\nabla}\cdot\bar{q} \right )+r,
\end{eqnarray}
where $\rho$ is the mass density,
$\hat{\sigma}_{\rm d}$ is the deviatoric component
of the Cauchy stress tensor $\hat{\sigma}$,
$\bar{q}$ is the heat flux vector,
$\Phi$ is the internal energy and $r$ is the heat supply per unit mass.
The Clausius--Duhem inequality implies that \cite{CG67}
\begin{equation}
\rho\frac{dQ}{dt}=\rho\frac{dS}{dt}
+\bar{\nabla} \cdot \Bigl (\frac{\bar{q}}{T}\Bigr )
-\frac{\rho r}{T} \geq 0,
\end{equation}
where $S$ is the entropy and $Q$ is the entropy production per unit mass.
Bearing in mind that
\[ \bar{\nabla}\cdot\Bigl (\frac{\bar{q}}{T}\Bigr )
=\frac{1}{T}\bar{\nabla}\cdot\bar{q}-\frac{1}{T^{2}}
\bar{q}\cdot\bar{\nabla} T, \]
and excluding the term $\bar{\nabla}\cdot\bar{q}$ from Eqs. (31)
and (32), we find that
\begin{equation}
T \frac{dQ}{dt}=T\frac{dS}{dt}-\frac{d\Phi}{dt}
+\frac{1}{\rho}\Bigl (\hat{\sigma}_{\rm d}:\hat{D}-\frac{1}{T}\bar{q}\cdot
\bar{\nabla} T \Bigr ) \geq 0.
\end{equation}
The internal energy $\Phi$ is given by the conventional formula
\[ \Phi=\Psi+ST, \]
where $\Psi$ is the free (Helmholtz) energy per unit mass.
Substitution of this equality into Eq. (33) yields
\begin{equation}
T \frac{dQ}{dt}=-S\frac{dT}{dt}-\frac{d\Psi}{dt}
+\frac{1}{\rho}\Bigl ( \hat{\sigma}_{\rm d}:\hat{D}-\frac{1}{T}\bar{q}\cdot
\bar{\nabla} T \Bigr) \geq 0.
\end{equation}
The following expression is accepted for the function $\Psi$:
\begin{eqnarray}
\Psi=\Psi_{0}+(c-S_{0})(T-T_{0}) -cT\ln\frac{T}{T_{0}} +U,
\end{eqnarray}
where $c$ is a specific heat,
and $\Psi_{0}$ and $S_{0}$ are the free energy
and the entropy per unit mass in the stress-free configuration
at the reference temperature $T_{0}$.
Formula (35) means that the configurational entropy (associated
with stretching of long chains) is neglected compared to mechanical
energy and only the entropy induced by thermal motion
\begin{equation}
S=S_{0}+c\ln\frac{T}{T_{0}}
\end{equation}
is taken into account.
The use of this hypothesis may be explained by the difficulties in
calculating the increment of the configurational entropy
driven by stretching of long chains \cite{AD90}, as well as
by some ambivalence in the definition of the mechanical energy, $U$,
for nonlinear entropic springs \cite{Eve95}.
The latter means that one cannot distinguish contributions of
the configurational entropy for non-Gaussian chains
and their mechanical energy into Eq. (35)
for the free energy per unit mass, unless physical assumptions
are specified for the choice of the function $W$ in Eq. (21).

Substitution of expressions (29), (35) and (36) into Eq. (34) results in
\begin{eqnarray}
T \frac{dQ}{dt} &=& J_{2}(t)
-\frac{1}{\rho T(t)}\bar{q}(t)\cdot\bar{\nabla}(t) T(t)
\nonumber\\
&& +\frac{1}{\rho}\biggl \{ \hat{\sigma}_{\rm d}
-2\rho \sum_{n=1}^{\infty} \mu(n)
\biggl [ X(t,0,n)\Bigl ( \psi_{01}(t)\hat{F}_{0}(t)
+\psi_{02}(t)\hat{F}_{0}^{2}(t)\Bigr )
\nonumber\\
&& +\int_{0}^{t} \frac{\partial X}{\partial \tau}(t,\tau,n)
\Bigl ( \psi_{1}(t,\tau)\hat{F}(t,\tau)
+\psi_{2}(t,\tau)\hat{F}^{2}(t,\tau)\Bigr ) d\tau\biggr ]
\biggr \}: \hat{D}(t).
\end{eqnarray}
It follows from Eqs. (24) and (37) that for an arbitrary loading program,
the rate of entropy production is nonnegative, provided that
\begin{enumerate}
\item
the heat flux vector $\bar{q}$ obeys the Fourier law
\[ \bar{q}(t)=-\kappa \bar{\nabla}(t)T(t) \]
with a positive thermal diffusivity $\kappa$,

\item
the Cauchy stress tensor $\hat{\sigma}$ satisfies the constitutive
equation
\begin{eqnarray}
\hat{\sigma}(t) &=& -P(t)\hat{I}+ 2\rho \sum_{n=1}^{\infty} \mu(n)
\biggl [ X(t,0,n)\Bigl ( \psi_{01}(t)\hat{F}_{0}(t)
+\psi_{02}(t)\hat{F}_{0}^{2}(t)\Bigr )
\nonumber\\
&& +\int_{0}^{t} \frac{\partial X}{\partial \tau}(t,\tau,n)
\Bigl ( \psi_{1}(t,\tau)\hat{F}(t,\tau)
+\psi_{2}(t,\tau)\hat{F}^{2}(t,\tau)\Bigr ) d\tau\biggr ],
\end{eqnarray}
where $P(t)$ is pressure.
\end{enumerate}
To employ Eq. (38) as a constitutive equation for a rubbery polymer,
one should establish connections between the relative Finger tensor
for a temporary network, $\hat{F}(t,\tau)$, and that for the polymer
at the macro-level.
The simplest assumption regarding these tensors is the affinity
hypothesis which postulates that $\hat{F}$ coincides
with the Finger tensor for macro-strains in a specimen.
A conventional explanation for this statement is that ``surrounding
molecules suppress the movements of the crosslinks so strongly that
their positions change affinely with the shape of the specimen''
\cite{Eve95}.

For a permanent network, where the rate of breakage and reformation of
chains vanish, Eqs. (4) and (38) are transformed into
the Finger formula for the Cauchy stress tensor
\[ \hat{\sigma} = -P\hat{I}+ 2\Lambda
\Bigl ( \psi_{01}\hat{F}_{0}+\psi_{02}\hat{F}_{0}^{2}\Bigr ) \]
with
\begin{equation}
\Lambda= \rho \Xi \sum_{n=1}^{\infty} \mu(n)p(n).
\end{equation}

\section{Uniaxial tension of a specimen}

In this Section, Eq. (38) is employed to determine stresses in a sample
at uniaxial extension.
Points of a polymeric specimen refer to a Cartesian frame $\{ X_{i} \}$
in the stress-free state and to a Cartesian frame $\{ x_{i} \}$
in the deformed state, $(i=1,2,3)$.
Uniaxial tension of an incompressible medium is described by the formulas
\[ x_{1}=k(t)X_{1},
\qquad
x_{2}=\biggl (\frac{1}{k(t)}\biggr )^{\frac{1}{2}} X_{2},
\qquad
x_{3}=\biggl (\frac{1}{k(t)}\biggr )^{\frac{1}{2}} X_{3}, \]
where $k$ is the extension ratio.
The relative deformation gradient $\bar{\nabla}(\tau)\bar{r}(t)$ reads
\[ \bar{\nabla}(\tau)\bar{r}(t)=\frac{k(t)}{k(\tau)}\bar{e}_{1}\bar{e}_{1}
+\biggl (\frac{k(\tau)}{k(t)}\biggr )^{\frac{1}{2}}
\Bigl (\bar{e}_{2}\bar{e}_{2}+\bar{e}_{3}\bar{e}_{3}\Bigr ), \]
where $\bar{e}_{i}$ are base vectors of the frame $\{ X_{i} \}$.
The relative Finger tensor $\hat{F}(t,\tau)$ is given by
\begin{equation}
\hat{F}(t,\tau)=\biggl (\frac{k(t)}{k(\tau)}\biggr )^{2}
\bar{e}_{1}\bar{e}_{1}
+\frac{k(\tau)}{k(t)}
\Bigl (\bar{e}_{2}\bar{e}_{2}+\bar{e}_{3}\bar{e}_{3}\Bigr ).
\end{equation}
Substituting Eq. (40) into Eq. (38), we find the non-zero components
\begin{eqnarray}
\sigma_{1}(t) &=& -P(t)+ 2\rho \sum_{n=1}^{\infty} \mu(n)
\biggl [ X(t,0,n) \Bigl ( \psi_{01}(t) k^{2}(t) +\psi_{02}(t) k^{4}(t)\Bigr )
\nonumber\\
&& +\int_{0}^{t} \frac{\partial X}{\partial \tau}(t,\tau,n)
\Bigl ( \psi_{1}(t,\tau)\biggl (\frac{k(t)}{k(\tau)}\biggr )^{2}
+\psi_{2}(t,\tau)\biggl ( \frac{k(t)}{k(\tau)}\biggr )^{4}
\Bigr ) d\tau\biggr ],
\nonumber\\
\sigma_{2}(t) &=& -P(t)+ 2\rho \sum_{n=1}^{\infty} \mu(n)
\biggl [ X(t,0,n) \Bigl ( \psi_{01}(t) k^{-1}(t)
+\psi_{02}(t) k^{-2}(t)\Bigr )
\nonumber\\
&& +\int_{0}^{t} \frac{\partial X}{\partial \tau}(t,\tau,n)
\Bigl ( \psi_{1}(t,\tau)\frac{k(\tau)}{k(t)}
+\psi_{2}(t,\tau)\biggl ( \frac{k(\tau)}{k(t)}\biggr )^{2}
\Bigr ) d\tau\biggr ]
\end{eqnarray}
of the Cauchy stress tensor
\[ \hat{\sigma}=\sigma_{1}\bar{e}_{1}\bar{e}_{1}
+\sigma_{2} \Bigl (\bar{e}_{2}\bar{e}_{2}+\bar{e}_{3}\bar{e}_{3}\Bigr ). \]
Excluding the unknown pressure $P(t)$ from Eq. (41) and the condition
\[ \sigma_{2}(t)=0, \]
we arrive at the formula for the longitudinal stress
\begin{eqnarray}
\sigma_{1}(t) &=& 2\rho \sum_{n=1}^{\infty} \mu(n)
\biggl \{ X(t,0,n) \Bigl [ \psi_{01}(t) \Bigl (k^{2}(t)-k^{-1}(t)\Bigr )
+\psi_{02}(t) \Bigl (k^{4}(t)-k^{-2}(t)\Bigr ) \Bigr ]
\nonumber\\
&& +\int_{0}^{t} \frac{\partial X}{\partial \tau}(t,\tau,n)
\Bigl [ \psi_{1}(t,\tau)\biggl (
\biggl (\frac{k(t)}{k(\tau)}\biggr )^{2}
-\frac{k(\tau)}{k(t)}\biggr )
\nonumber\\
&& +\psi_{2}(t,\tau)\biggl (
\biggl ( \frac{k(t)}{k(\tau)}\biggr )^{4}
-\biggl ( \frac{k(\tau)}{k(t)}\biggr )^{2}\biggr )
\Bigr ] d\tau\biggr \},
\end{eqnarray}
where the function $X$ is determined by Eqs. (4) and (9) and
the functions $\psi_{k}$ are given by Eq. (30).

We confine ourselves to the Mooney--Rivlin material
with the mechanical energy \cite{Moo40,RS51}
\begin{equation}
W(I_{1},I_{2})=c_{1}(I_{1}-3)+c_{2}(I_{2}-3),
\end{equation}
where $c_{1}$ and $c_{2}$ are adjustable parameters.
The choice of the Mooney--Rivlin equation may be explained by two reasons:
(i) formula (43) fits observations for some rubbery polymers
and polymeric melts with a high level of accuracy \cite{ZMF86}
and (ii) Eq. (43) provides a compromise between the concepts of
affine and phantom  networks \cite{NR94}.

Substitution of Eqs. (30) and (43) into Eq. (42) results in
\begin{eqnarray*}
\sigma_{1}(t) &=& 2\rho \sum_{n=1}^{\infty} \mu(n)
\biggl [ X(t,0,n) \Bigl ( c_{1}+c_{2} k^{-1}(t)\Bigr )
\Bigl (k^{2}(t)-k^{-1}(t)\Bigr )
\nonumber\\
&& +\int_{0}^{t} \frac{\partial X}{\partial \tau}(t,\tau,n)
\Bigl ( c_{1}+c_{2}\frac{k(\tau)}{k(t)}\Bigr )
\biggl (\Bigl (\frac{k(t)}{k(\tau)}\Bigr )^{2}
-\frac{k(\tau)}{k(t)}\biggr ) d\tau\biggr ].
\end{eqnarray*}
Combining this equality with Eqs. (4) and (9), we arrive at
the stress--strain relation
\begin{eqnarray}
\sigma_{1}(t) &=& 2\rho \Xi \sum_{n=1}^{\infty} \mu(n)p(n)
\biggl [ \exp \Bigl (-\int_{0}^{t} \Gamma(s,n)ds\Bigr )
\Bigl ( c_{1}+c_{2} k^{-1}(t)\Bigr ) \Bigl (k^{2}(t)-k^{-1}(t)\Bigr )
\nonumber\\
&& +\int_{0}^{t} \Gamma(\tau,n)
\exp\Bigl (-\int_{\tau}^{t} \Gamma(s,n) ds\Bigr )
\Bigl ( c_{1}+c_{2}\frac{k(\tau)}{k(t)}\Bigr )
\biggl (\Bigl (\frac{k(t)}{k(\tau)}\Bigr )^{2}
-\frac{k(\tau)}{k(t)}\biggr ) d\tau\biggr ].
\end{eqnarray}
For a permanent network, Eq. (44) is reduced to the conventional formula
\begin{equation}
\sigma_{1}=2(C_{1}+C_{2}k^{-1})(k^{2}-k^{-1})
\end{equation}
with
\begin{equation}
C_{i}=\Lambda c_{i}.
\end{equation}
We now analyze the viscoelastic response of a rubbery polymer,
when small oscillations with a fixed amplitude $k_{1}$
and a fixed frequency $\omega$ are superposed on the time-independent
stretching of a specimen with the longitudinal elongation $k_{0}$,
\begin{equation}
k(t)=k_{0}+k_{1}\exp ({\rm i}\omega t).
\end{equation}
Assuming $k_{0}$ to be of the order of unity and
$k_{1}$ to be small compared to unity and neglecting terms of
the second order of smallness, we find that
\begin{eqnarray*}
c_{1}+c_{2}\frac{k(\tau)}{k(t)} &=& c_{1}+c_{2}\biggl [1-
\frac{k_{1}}{k_{0}}\Bigl (\exp ({\rm i}\omega t)-\exp ({\rm i}\omega \tau)
\Bigr )\biggr ],
\nonumber\\
c_{1}+c_{2} k^{-1}(t) &=& c_{1}+c_{2} k_{0}^{-1}
-\frac{k_{1}}{k_{0}^{2}} c_{2} \exp ({\rm i}\omega t),
\nonumber\\
\Bigl (\frac{k(t)}{k(\tau)}\Bigr )^{2}-\frac{k(\tau)}{k(t)}
&=& 3\frac{k_{1}}{k_{0}}\Bigl (\exp ({\rm i}\omega t)-\exp ({\rm i}\omega \tau)
\Bigr ),
\nonumber\\
k^{2}(t)-k^{-1}(t) &=& (k_{0}^{2}-k_{0}^{-1})
+\frac{k_{1}}{k_{0}}(2 k_{0}^{2}+k_{0}^{-1})\exp ({\rm i}\omega t).
\end{eqnarray*}
It follows from these equalities that with the required level of
accuracy,
\begin{eqnarray}
&& \Bigl (c_{1}+c_{2}k^{-1}(t)\Bigr )\Bigl (k^{2}(t)-k^{-1}(t)\Bigr )
= (c_{1}+c_{2}k_{0}^{-1})(k_{0}^{2}-k_{0}^{-1})
\nonumber\\
&& +\frac{k_{1}}{k_{0}}\biggl [ c_{1}(2k_{0}^{2}+k_{0}^{-1})
+c_{2}k_{0}^{-1}(k_{0}^{2}+2k_{0}^{-1})\biggr ]\exp ({\rm i}\omega t),
\nonumber\\
&& \Bigl (c_{1}+c_{2}\frac{k(\tau)}{k(t)}\Bigr )
\biggl (\Bigl (\frac{k(t)}{k(\tau)}\Bigr )^{2}-\frac{k(\tau)}{k(t)}\biggr )
=3\frac{k_{1}}{k_{0}}(c_{1}+c_{2})\Bigl (\exp ({\rm i}\omega t)
-\exp ({\rm i}\omega \tau)\Bigr ).
\end{eqnarray}
Substitution of Eq. (48) into Eq. (44) implies that
\[ \sigma_{1}(t)=S_{0}(t)+S_{1}(t)+S_{2}(t), \]
where
\[ S_{0}(t)=2\rho (c_{1}+c_{2}k_{0}^{-1})
(k_{0}^{2}-k_{0}^{-1})\sum_{n=1}^{\infty} \mu(n)X(t,0,n) \]
determines relaxation of stresses in a specimen stretched with
the extension ratio $k_{0}$,
\begin{eqnarray*}
S_{1}(t) &=& 2\rho \frac{k_{1}}{k_{0}}
\biggl [ \Bigl (c_{1}(2k_{0}^{2}+k_{0}^{-1})
+c_{2}k_{0}^{-1}(k_{0}^{2}+2k_{0}^{-1})\Bigr )-3\biggr ]
\nonumber\\
&&\times \exp ({\rm i}\omega t)
\sum_{n=1}^{\infty} \mu(n) X(t,0,n)
\end{eqnarray*}
characterizes relaxation of small oscillatory stresses and
\begin{eqnarray*}
S_{2}(t) &=& 6\rho \frac{k_{1}}{k_{0}}(c_{1}+c_{2})
\sum_{n=1}^{\infty} \mu(n) \biggl [ X(t,t,n) \exp ({\rm i}\omega t)
\nonumber\\
&& -\int_{0}^{t} \frac{\partial X}{\partial \tau}(t,\tau,n)
\exp ({\rm i}\omega \tau)d\tau \biggr ]
\nonumber\\
&=& 6\rho \frac{k_{1}}{k_{0}}(c_{1}+c_{2})\Xi
\sum_{n=1}^{\infty} \mu(n) p(n) \biggl [ 1
\nonumber\\
&& -\int_{0}^{t} \Gamma(\tau,n) \exp \biggl (-\int_{\tau}^{t}
\Gamma(s,n)ds -{\rm i}\omega (t-\tau)\biggr ) d\tau \biggr ]
\exp ({\rm i}\omega t)
\end{eqnarray*}
determines the steady response to small oscillations
superposed on uniaxial stretching.
Following \cite{Lio98}, we define the complex elastic modulus
$E^{\ast}$ as the ratio of $S_{2}(t)$ to the
oscillatory strain
\[ \Delta \epsilon(t)=\frac{k_{1}}{k_{0}} \exp({\rm i}\omega t) \]
and obtain
\begin{eqnarray}
E^{\ast}(t,\omega) &=& 6\rho (c_{1}+c_{2})\Xi
\sum_{n=1}^{\infty} \mu(n) p(n) \biggl [ 1
\nonumber\\
&& -\int_{0}^{t} \Gamma(\tau,n) \exp \biggl (-\int_{\tau}^{t}
\Gamma(s,n)ds -{\rm i}\omega (t-\tau)\biggr ) d\tau \biggr ].
\end{eqnarray}
For the loading process (47),
the rate of breakage $\Gamma$ is time-independent, $\Gamma=\Gamma(n)$,
(it is determined by the time-independent ``basic'' elongation $k_{0}$).
Introducing the new variable $s=t-\tau$ and
replacing the upper limit of integration in Eq. (49) by infinity
(in agreement with the conventional procedure for the analysis
of dynamic response), we obtain
\begin{eqnarray*}
E^{\ast}(\omega) &=& 6\rho (c_{1}+c_{2})\Xi
\sum_{n=1}^{\infty} \mu(n) p(n) \biggl [ 1
-\Gamma(n) \int_{0}^{\infty} \exp \Bigl (-\Bigl (\Gamma(n)+
{\rm i}\omega \Bigr ) (t-\tau)\Bigr ) d\tau \biggr ]
\nonumber\\
&=& 6\rho (c_{1}+c_{2})\Xi \sum_{n=1}^{\infty} \mu(n) p(n)
\Bigl [ 1-\frac{\Gamma(n)}{\Gamma(n)+{\rm i}\omega}\Bigr ].
\end{eqnarray*}
We split the complex modulus $E^{\ast}$ into the sum of
storage and loss moduli,
\[ E^{\ast}(\omega)=E^{\prime}(\omega)+{\rm i}E^{\prime\prime}(\omega), \]
and arrive at the formulas
\begin{eqnarray}
E^{\prime}(\omega) &=& 6\rho (c_{1}+c_{2})\Xi \sum_{n=1}^{\infty}
\frac{\mu(n) p(n)\omega^{2}}{\Gamma^{2}(n)+\omega^{2}},
\nonumber\\
E^{\prime\prime}(\omega) &=& 6\rho (c_{1}+c_{2})\Xi \sum_{n=1}^{\infty}
\frac{\mu(n) p(n)\Gamma(n) \omega}{\Gamma^{2}(n)+\omega^{2}}.
\end{eqnarray}
It follows from Eq. (50) that for a Mooney--Rivlin viscoelastic
medium, the dynamic moduli are independent of the elongation, $k_{0}$,
on which small oscillations are superposed.
This result is a substantial shortcoming of Eq. (43) because experimental
data reveal that the storage and loss moduli noticeable change
with stretching.

\section{Adjustable parameters in the model}

Governing equations (45) and (50) are determined by two adjustable
parameters, $c_{1}$ and $c_{2}$, which characterize the elastic
response of a polymer at finite strains, and three material functions,
$p(n)$, $\mu(n)$ and $\Gamma(t,n)$.
The function $p(n)$ describes the distribution of long chains with
various numbers of strands.
This function is analogous to the probability density of traps
with various energies in the energy-landscape theory for structural
glasses \cite{Dyr95}.
Because the distribution of long chains in rubbery polymers is determined
by the vulcanization process (whose description at the micro-level is far
from being exhausted, see \cite{GCZ96} and the references therein),
no explicit expression has been yet proposed for this function.
As a first approximation, we assume the distribution of chains
to be exponential
\begin{equation}
p(n)=\frac{\exp(-\alpha n)}{\exp(\alpha)-1},
\end{equation}
where $\alpha$ is an positive constant.
The factor $\exp(\alpha)-1$ in Eq. (51) is determined from the condition
\[ \sum_{n=1}^{\infty} p(n)=1 \]
and the formula
\begin{equation}
\sum_{n=1}^{\infty} \exp (-\alpha n)= \sum_{n=0}^{\infty} \Bigl [
\exp (-\alpha)\Bigr ]^{n}-1=\frac{1}{1-\exp (-\alpha)}-1
=\frac{1}{\exp (\alpha)-1}.
\end{equation}
The average number of strands in a chain is given by
\[ \langle n\rangle =\sum_{n=1}^{\infty} n p(n)=\frac{1}{\exp(\alpha)-1}
\sum_{n=1}^{\infty} n\exp (-\alpha n)
=-\frac{1}{\exp(\alpha)-1}\frac{\partial}{\partial \alpha}
\sum_{n=1}^{\infty} \exp (-\alpha n). \]
Combining this equality with Eq. (52), we find that
\begin{equation}
\langle n\rangle =\frac{\exp (\alpha)}{(\exp(\alpha)-1)^{3}}.
\end{equation}

The other adjustable function, $\mu(n)$, characterizes the
longitudinal rigidity of a long chain containing $n$ strands.
With reference to the Rouse model, a chain may be treated
as a system of $n$ identical springs connected in sequence.
The rigidity of such a system is given by the conventional formula
\begin{equation}
\mu(n)=\frac{\mu_{0}}{n},
\end{equation}
where $\mu_{0}$ is the rigidity of an individual spring.
We suppose that Eq. (54) may be applied (as a first approximation)
to long chains with an arbitrary geometry.

The third function to be found is the rate of relaxation $\Gamma(t,n)$.
It is assumed that $\Gamma(t,n)$ is factorized as
\begin{equation}
\Gamma(t,n)=\Gamma_{0}(t)\eta(n),
\end{equation}
where $\Gamma_{0}$ is responsible for mechanically-induced changes
in the relaxation rate,
whereas $\eta(n)$ describes the effect of the chain's length
on the rate of its slippage from a temporary junction.
We suppose that $\eta$ exponentially grows with the number of strands $n$,
\begin{equation}
\eta(n)=\exp (\beta n),
\end{equation}
where $\beta$ is a positive parameter.
Equation (56) is in contrast with the Eyring formula which is
conventionally adopted for the description of thermally activated
processes in polymers \cite{KE75}.
Within the concept of traps, the rate of relaxation is traditionally
presumed to decrease exponentially with the energy of a potential well
on the energy landscape where a relaxing region is trapped.
Assuming this energy to be proportional to the volume of a micro-domain,
we find that the relaxation rate decreases with the number of strands
participating in collective rearrangement, whereas Eq. (56) implies
that this rate grows with the number of strands in a long chain.
This contradiction may be explained with reference to the reptation
theory for polymeric chains \cite{DE86}.
A semiflexible chain with a finite bending stiffness
is thought of as a curvilinear rod whose micro-motion
is restricted to a tube composed by surrounding chains.
The radius of the tube is estimated as a few lengths of a strand,
which implies that the micro-motion of a chain may be split
into diffusion along the tube and small lateral fluctuations
around the tube's centerline.
For a stiffless rod, lateral fluctuations do not affect the response
of junctions, which implies that breakage of chains is mainly associated
with longitudinal diffusion.
In this case, small thermal fluctuations induce rather small
longitudinal displacements whose energy is insufficient
to detach a chain with a large number of strands from the junctions.
The opposite picture is observed for a chain with a finite
bending stiffness, whose lateral oscillations (driven by relatively
small, but random fluctuations) may be amplified due to interaction of
transverse elastic waves in the rod and may cause tearing of
a chain from temporary junctions.
This study deals with long chains with a non-vanishing bending
stiffness, for which Eq. (56) serves as a natural hypothesis.

Substitution of expressions (51) and (54) into Eq. (39) results in
\[ \Lambda=\frac{\mu_{0} \rho\Xi}{\exp (\alpha)-1} \sum_{n=1}^{\infty}
\frac{\exp (-\alpha n)}{n}. \]
Bearing in mind that
\[ \frac{\exp(-\alpha n)}{n}=\int_{\alpha}^{\infty} \exp (-nx) dx, \]
we rewrite this equality as follows:
\[ \Lambda=\frac{\mu_{0} \rho\Xi}{\exp (\alpha)-1} \int_{\alpha}^{\infty}
\sum_{n=1}^{\infty} \exp (-\alpha x) dx
=\frac{\mu_{0} \rho\Xi}{\exp (\alpha)-1} \int_{\alpha}^{\infty}
\frac{dx}{\exp (x)-1}, \]
where the sum is determined with the help of Eq. (52).
To calculate the integral, we introduce the new variable $y=\exp(x)$
and obtain
\[ \int_{\alpha}^{\infty} \frac{dx}{\exp (x)-1}
=\int_{\exp(\alpha)}^{\infty} \frac{dy}{y(y-1)}
=\int_{\exp(\alpha)}^{\infty} \Bigl (\frac{1}{y-1}-\frac{1}{y}\Bigr )dy
=\alpha-\ln \Bigl (\exp(\alpha)-1\Bigr ). \]
This implies that
\begin{equation}
\Lambda=\mu_{0} \rho\Xi
\frac{\alpha-\ln (\exp(\alpha)-1)}{\exp (\alpha)-1}.
\end{equation}
Setting
\[ \Xi=\frac{\Xi_{0}}{\langle n\rangle}, \]
where $\Xi_{0}$ is the number of strands per unit mass,
we find that
\begin{equation}
\Lambda=\Lambda_{0}\frac{\alpha-\ln (\exp(\alpha)-1)}{\langle n\rangle
(\exp (\alpha)-1)}
\end{equation}
with $\Lambda_{0}=\mu_{0} \rho\Xi_{0}$.
Equations (53) and (58) allow the average rigidity of an ensemble
of chains, $\Lambda$, to be expressed as a function of
the average number of strands in a chain $\langle n\rangle$.
Figure~1 reveals that the quantity $\Lambda$ [an analog of
the elastic modulus of a rubbery polymer, see Eq. (46)]
decreases with the average number of strands in a chain $\langle n\rangle$.
With an acceptable level of accuracy this dependence may be described
by the scaling law
\[ \Lambda \propto \langle n\rangle^{-\kappa} \]
with $\kappa=0.56$.

\section{Validation of the model}

Our objective now is to find adjustable parameters of the model
by fitting observations for a carbon black filled rubber
at various temperatures, $T$, and various programs of loading.

\subsection{Tensile tests with high rates of loading}

We begin with matching observations in tensile tests
with relatively high rates of loading (when the processes of breakage
and reformation of long chains during the test may be neglected)
and moderate deformation (when mechanically induced alignment of chains
is not observed and the network may be treated as isotropic).
For a detailed description of specimens and the experimental procedure,
see \cite{Lio96}.

The loading process is characterized by the rate of engineering strain
\[ \dot{\epsilon}_{\rm eng}=\frac{d\epsilon_{\rm eng}}{dt},
\qquad
\epsilon_{\rm eng}=k-1. \]
Experimental data in tensile tests with the strain rate
$\dot{\epsilon}_{\rm eng}=0.2$ s$^{-1}$ at four different temperatures
are plotted in Figure~2 together with the results of numerical simulation.
The parameters $C_{1}$ and $C_{2}$ in Eq. (45) are determined using
the least-squares algorithm.

It is found that the quantity $C_{1}$ may be set to be zero,
and the model with only one adjustable parameter, $C_{2}$,
correctly predicts observations at temperatures from $T=296$~K to
$T=373$~K.
At the lowest temperature, $T=253$~K, experimental data slightly
deviate from the model prediction.
These discrepancies may be explained by (partial) failure of
carbon filled rubber and insufficient accuracy of the Mooney--Rivlin
equation (43).

The influence of temperature $T$ on the elastic modulus is
illustrated by Figure~3, where the quantity $C_{2}$ is plotted
versus the degree of undercooling $\Delta T=T-T_{\rm g}$.
The elastic modulus monotonically decreases with temperature
until some critical temperature $T_{\rm cr}$
and, afterwards, remains practically constant.
In the interval $[T_{\rm g}, T_{\rm cr}]$ the dependence $C_{2}(T)$
is fairly well approximated by the linear function
\begin{equation}
C_{2}=a_{0}-a_{1}\Delta T,
\end{equation}
where the quantities $a_{0}$ and $a_{1}$ are found using the
least-squares technique.
The decrease in $C_{2}$ with temperature confirms our hypothesis
that in the region of temperatures under consideration, the effect
of configurational entropy (whose contribution into the free energy
linearly increases with temperature) is negligible compared to the
mechanical energy of chains.

\subsection{Dynamical tests on a compressed specimen}

As it was discussed in Introduction, an increase in the elastic modulus
of a polymer with a decrease in temperature is traditionally
associated with two phenomena:
(i) an increase in the rigidity of a strand, $\mu_{0}$,
and (ii) an increase in the number of temporary junctions
(which is tantamount to a decrease in the average number of strands in
a chain, $\langle n\rangle$).
To assess which mechanism is responsible for temperature-dependent
changes in elastic moduli and to analyze the effect of temperature
on the rate of breakage of chains, we fit experimental data for
a preloaded specimen (the longitudinal strain $\epsilon_{0}=-0.1$)
in dynamic tests with the small amplitude
of oscillations $\Delta\epsilon=0.006$.
A detailed description of the experimental procedure can be found
in \cite{Lio98}.

Substitution of Eqs. (51), (54) and (56) into Eq. (50) results in
\begin{equation}
E^{\prime}(\omega)=C\sum_{n=1}^{\infty}
\frac{\omega^{2}\exp (-\alpha n)}{n(\Gamma_{0}\exp (\beta n)+\omega^{2})},
\end{equation}
where
\begin{equation}
C=\frac{6\pi \mu_{0} \rho \Xi}{\exp(\alpha)-1} (c_{1}+c_{2}).
\end{equation}
Comparing Eqs. (46) and (61) and bearing in mind Eq. (57) and the equality
$C_{1}=0$, we find that
\begin{equation}
C=\frac{6C_{2}}{\alpha-\ln(\exp(\alpha)-1)}.
\end{equation}
Equation (62) provides a simple relation between the moduli
observed in a uniaxial tensile test with a constant rate of strain
and in a dynamic test with a small amplitude of oscillations.
However, this formula should be taken with caution, because its
derivation is based on the assumption about the Mooney--Rivlin
equation for the strain energy density of a chain (the hypothesis
which does not guarantee an acceptable quality of fitting
observations in tensile tests with a constant rate of strain,
and which may lead to even larger discrepancies between the
linearized equation (50) and experimental data in dynamic tests).
It is easy to show that for any strain energy density of a chain, $W$,
[not necessarily determined by Eq. (43)],
the parameter $C$ is determined by Eq. (61),
where the sum $c_{1}+c_{2}$ is replaced by some
coefficient, which, in general, depends on the strain $\epsilon_{0}$,
but which is independent of temperature, $T$, and the average number
of strands, $\langle n\rangle$.

To check which of the two assumptions about the influence of temperature
is more adequate for the description of the response of rubbery polymers,
we suppose that the number of strands is temperature-independent
(which is tantamount to the postulate that $\alpha$ is independent
of temperature) and fit observations in dynamic tests at various
temperatures.
We set $\alpha=0.02$, which corresponds to the average number of strands
$\langle n\rangle=1.24\times 10^{5}$ (the value which is in agreement
with available experimental data for rubbery polymers \cite{Fer80}).
The parameter $\beta=2.27$ is found by fitting data in a test at ambient
temperature by using the steepest-descent procedure.
We fix this value of $\beta$ and approximate observations
at other temperatures with only two adjustable constants,
$C$ and $\Gamma_{0}$.
Given a parameter $C$, the relaxation rate $\Gamma_{0}$ is determined
using the steepest-descent algorithm.
An analog of rigidity, $C$, is found by the
least-squares technique.

Figure~4 demonstrates good agreement between experimental data and
results of numerical simulation with a time-independent parameter $\alpha$.
This means that our assumption that the number of strands remains
constant at all temperatures in the region under consideration
is quite acceptable for matching observations.

The parameter $C$ is plotted versus temperature in Figure~3
which demonstrates that in the region of temperatures
$[T_{\rm g},T_{\rm cr}]$, the dependence $C(T)$ is fairly well
approximated by the linear function
\begin{equation}
C=b_{0}-b_{1}\Delta T,
\end{equation}
where the parameters $b_{0}$ and $b_{1}$ are determined by the
least-squares technique.
Because the slopes of the graphs $C_{2}(\Delta T)$ and $C(\Delta T)$
are close to each other (the ratio $a_{1}/a_{0}$ is 0.0071,
whereas $b_{1}/b_{0}$ equals 0.0067),
we may conclude that in the interval $[T_{\rm g},T_{\rm cr}]$,
the effect of temperature on $\alpha$ is rather weak.
This means that the average number of strands in a chain,
$\langle n\rangle$, feebly depends on temperature,
which implies that entanglements are practically not
transformed into temporary junctions.

The relaxation rate $\Gamma_{0}$ is plotted versus the degree of
undercooling $\Delta T$ in Figure~5.
Experimental data are fairly well approximated by the function
\begin{equation}
\log \Gamma_{0}=d_{0}-d_{1}\Delta T,
\end{equation}
where adjustable parameters $d_{0}$ and $d_{1}$
are found using the least-squares algorithm.

Equation (64) implies that the rate of breakage for long chains
decreases with temperature (in contrast with the theory of thermally
activated processes which predicts the growth of $\Gamma_{0}$).
This result may be explained by a decrease in the bending
stiffness of chains with temperature.
At low temperatures, when the bending rigidity is relatively high,
any local thermal fluctuation produces oscillations
of a chain (which is thought of as a curvilinear elastic rod).
Interaction of transverse oscillations driven at random times
at random points of the rod with fixed ends amplifies their amplitude,
and, as a consequence, induces relatively large displacements
of the chain's ends
(local thermal fluctuations are transformed into global ones
at the length-scale of a chain).
The amplification of random displacements of the chain's ends
may be sufficient for their slippage from temporary junctions,
which is reflected in the high rate of breakage, $\Gamma_{0}$,
in the vicinity of the glass transition point.
With the growth of temperature, the bending rigidity substantially
decreases, which results in a decline of transverse oscillations
of chains.
Local thermal fluctuations weakly interact and their amplitudes
are not amplified.
As a result, only thermal fluctuations in the close vicinity of
a chain's ends can induce their slippage from temporary junctions.
Although the number of these fluctuations increases
and the strength of junctions decreases with temperature,
the total number of broken chains (per unit time)
diminishes, in agreement with data depicted in Figure~5.

\section{Concluding remarks}

Constitutive equations are derived for the nonlinear viscoelastic
response of rubbery polymers at finite strains.
The model is based on the concept of temporary networks, which treats
an amorphous polymer as an ensemble of long chains connected to
junctions.
At random times, active chains detach from the junctions as they are
thermally agitated.
A dangling chain merges with the network when its free end captures
some junction in its vicinity.

Unlike conventional concepts of transient networks, we suppose that (i)
long chains consist of various numbers of strands and (ii)
the rigidity of a chain and the rate of its breakage substantially
depend on the chain's length.
Several hypotheses are introduced regarding the distribution of chains
with various lengths and the dependence of the rate of reformation
on the number of strands.
These assertions are verified by comparison with observations in
uniaxial tensile tests.
Fair agreement is demonstrated between experimental data for a carbon
black filled rubber at various temperatures and results of numerical
simulation.

The following conclusions are drawn:
\begin{enumerate}
\item
The average number of strands in a chain weakly depends on temperature.
This implies that changes in the viscoelastic response of rubber
with temperature may be ascribed to the effect of
temperature on the average rigidity of a strand only.
An acceptable agreement with experimental data is achieved
without the hypothesis about a substantial increase in the number of
junction with a decrease in temperature (induced by transition
of entanglements into temporary junctions).

\item
Some critical temperature, $T_{\rm cr}$, is found
above the glass transition point $T_{\rm g}$.
In the region $[T_{\rm g},T_{\rm cr}]$, the rigidity
of a strand linearly decreases with temperature, while above the
critical point, this parameter is independent of temperature.
The values of $T_{\rm cr}$ determined in static and dynamic tests
are close to one another.

\item
Conventional theories of rubber elasticity are based on the assumption
about the entropic nature of free energy.
This implies that stresses linearly increase with temperature.
Because this assertion contradicts experimental data for carbon black
filled rubber, it is postulated that the entropic contribution
into the free energy is small compared to that for the mechanical energy.
This hypothesis is fairly well confirmed by observations, which
demonstrate a decrease in stresses with the growth of temperature.

\item
Unlike the theory of cooperative relaxation in glassy polymers
which presumes that the rate of rearrangement exponentially
decreases with volume of a rearranged domain,
the rate of breakage for rubbery polymers increases with the growth
of the number of strands in a long chain.
This may be explained by a finite bending rigidity of a chain which
is not taken into account by conventional theories.
At relatively large bending rigidity, an active chain may
be thought of as a curvilinear elastic rod.
Local thermal fluctuations result in transverse oscillations of the rod.
The amplitude of oscillations grows because of their interaction,
which results in an increase in transverse displacements.
The growth of the bending rigidity and the rod's length
leads to an increase in the amplitude of oscillations for
the rod's ends, and, as a consequence, to an increase in
the rate of breakage.

\item
In contrast with the theory of thermally activated processes,
the rate of breakage for active chains decreases with temperature.
This may be explained by a substantial decline in the bending rigidity
of chains which implies that local random oscillations do not
interact, and only thermal fluctuations in the close vicinity of
the end points cause detachment of an active chain from the network.
\end{enumerate}

\subsection*{Acknowledgement}

AD gratefully acknowledge financial support by the Israel Ministry
of Science through grant 1202-1-00.

\newpage

\setlength{\unitlength}{1.0 mm}
\begin{figure}[t]
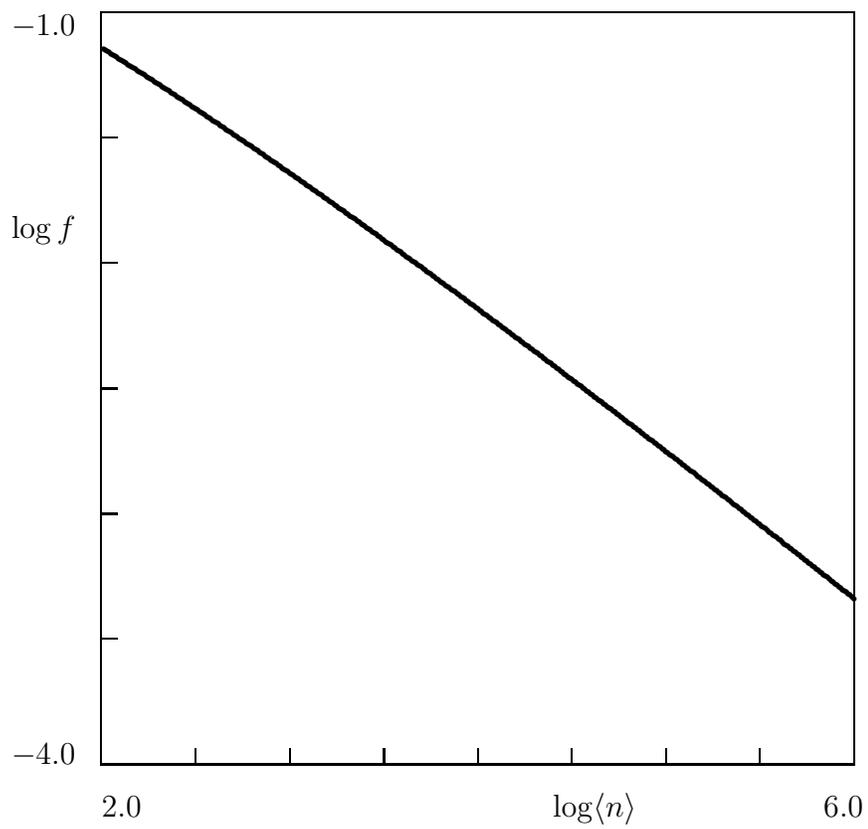

\begin{center}

\end{center}
\vspace*{10 mm}

\caption{The ratio $f=\Lambda/\Lambda_{0}$
versus the average number of strands in a chain $\langle n\rangle$}
\end{figure}

\begin{figure}[t]
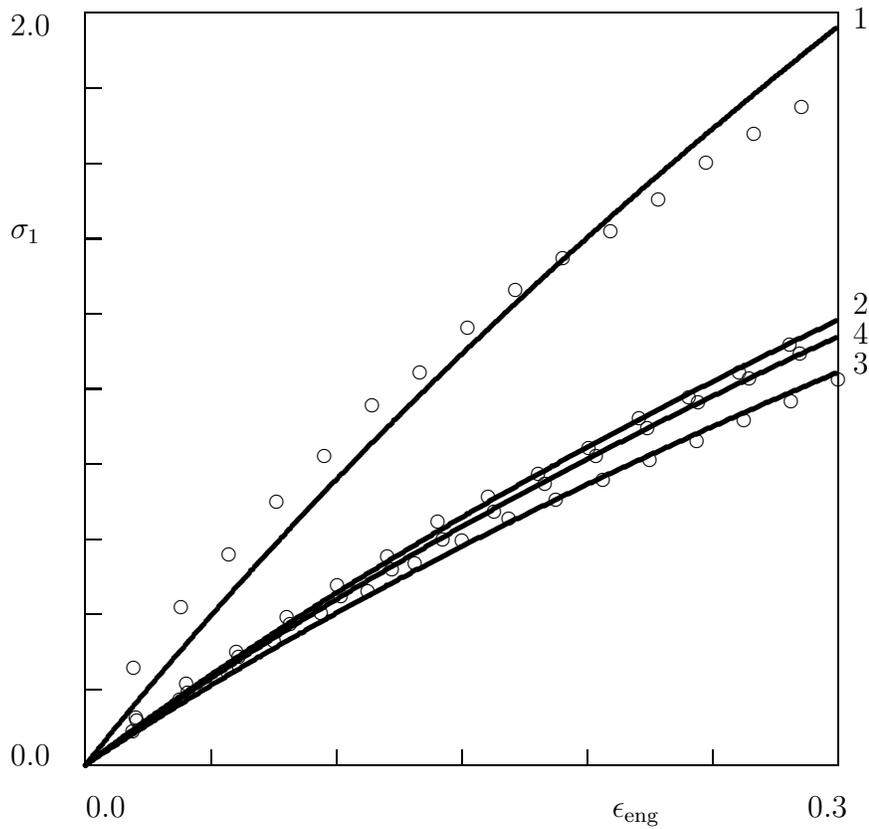

\begin{center}
%
\end{center}
\vspace*{10 mm}

\caption{The longitudinal stress $\sigma_{1}$ MPa versus the engineering
strain $\epsilon_{\rm eng}$ for carbon black filled rubber
in tensile tests at various temperatures $T$~K.
Circles: experimental data \protect{\cite{Lio97b}}.
Solid lines: their approximation by Eq. (45).
Curve~1: $T=253$;
curve~2: $T=296$;
curve~3: $T=333$;
curve~4: $T=373$}
\end{figure}

\begin{figure}[t]
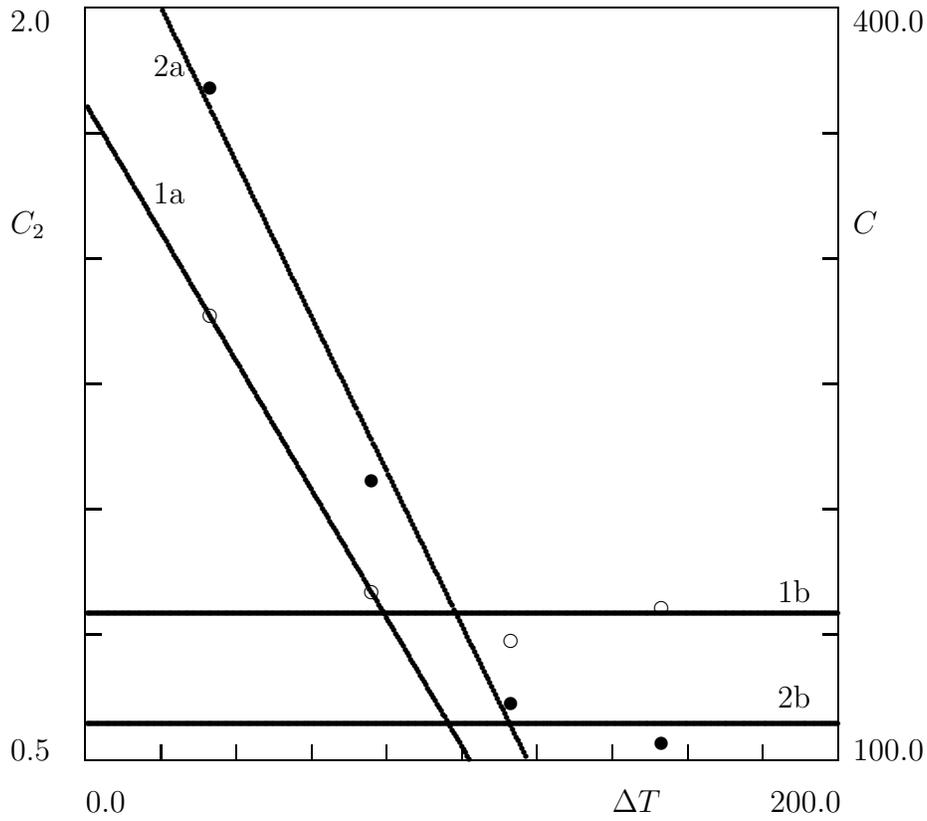

\begin{center}
%
\end{center}
\vspace*{10 mm}

\caption{The coefficients $C_{2}$ MPa (unfilled circles)
and $C$ MPa (filled circles) versus the degree of undercooling
$\Delta T$~K for a carbon black filled rubber.
Circles: treatment of observations \protect{\cite{Lio97b,Lio98}}.
Solid lines: approximation of the experimental data by Eqs. (59) and 63).
Curve~1a: $a_{0}=1.8097$, $a_{1}=0.0128$;
curve~2a: $b_{0}=462.16$, $b_{1}=3.0835$}
\end{figure}

\begin{figure}[t]
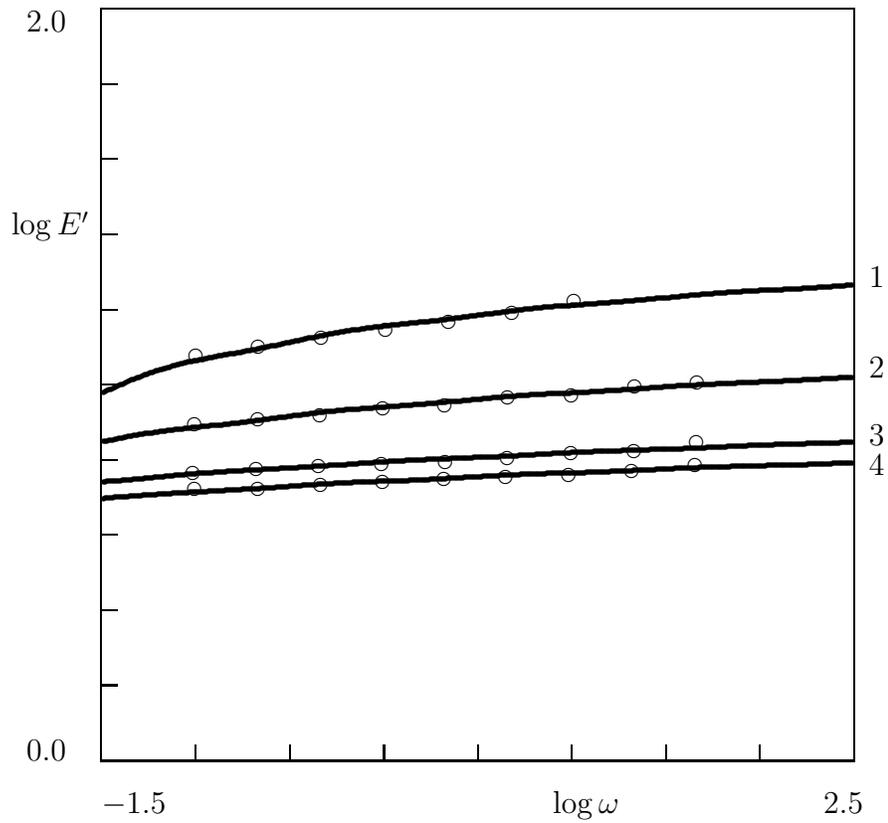

\begin{center}
%
\end{center}
\vspace*{10 mm}

\caption{The storage modulus $E^{\prime}$ MPa versus
the frequency of oscillations $\omega$ Hz for carbon black
filled rubber in tensile dynamic tests on preloaded specimens
($\epsilon_{0}=-0.1$) with the amplitude of oscillations
$\Delta\epsilon=0.006$ at temperature $T$~K.
Circles: experimental data \protect{\cite{Lio98}}.
Solids lines: results of numerical simulation.
Curve~1: $T=253$;
curve~2: $T=296$;
curve~3: $T=333$;
curve~4: $T=373$}
\end{figure}

\begin{figure}[t]
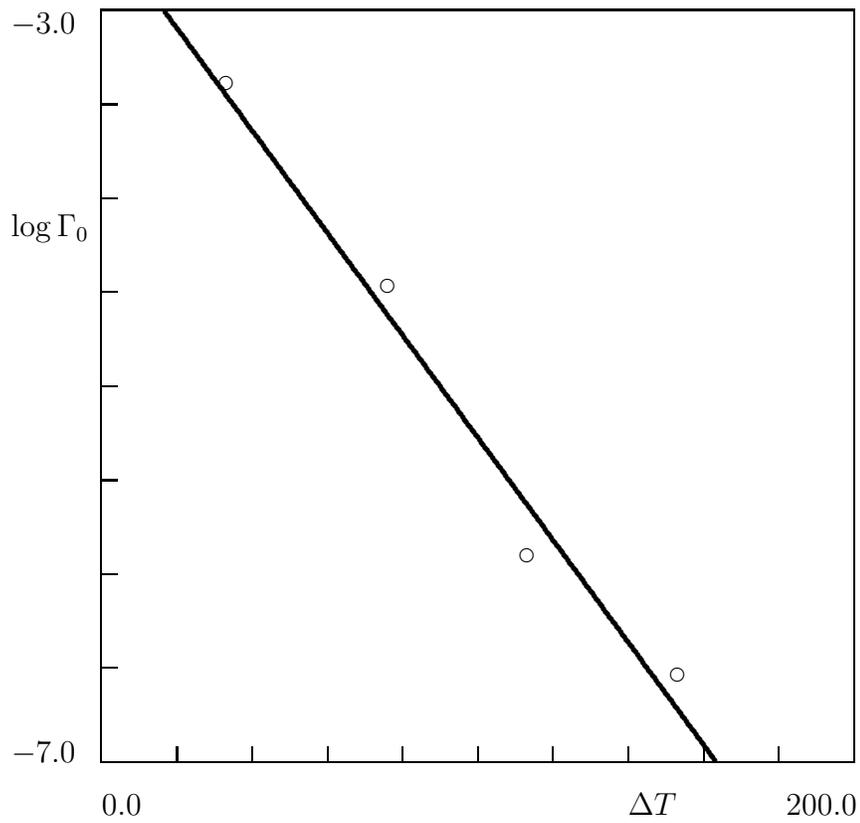

\begin{center}
%
\end{center}
\vspace*{10 mm}

\caption{The rate of breakage $\Gamma_{0}$ s$^{-1}$
versus the degree of undercooling $\Delta T$~K
for a carbon black filled rubber.
Circles: treatment of observations \protect{\cite{Lio98}}.
Solid lines: approximation of the experimental data by Eq. (64)
with $d_{0}=-2.5489$ and $d_{1}=0.0272$}
\end{figure}

\newpage
\begin{thebibliography}{80}
\bibitem{GS91}
Govinjee, S., Simo, J.:
A micro-mechanically based continuum damage model for carbon black-filled
rubbers incorporating Mullins effect.
{\em J. Mech. Phys. Solids} {\bf 39}, 87--112 (1991).

\bibitem{SC91}
So, H., Chen, U.D.:
A nonlinear mechanical model for solid-filled rubbers.
{\em Polym. Eng. Sci.} {\bf 31}, 410--416 (1991).

\bibitem{GS92}
Govinjee, S., Simo, J.:
Mullins' effect and the strain amplitude dependence of the storage
modulus.
{\em Int. J. Solids Structures} {\bf 29}, 1737--1751 (1992).


\bibitem{AB93}
Arruda, E.M., Boyce, M.C.:
A three dimensional constitutive model for the large stretch
behavior of rubber elastic materials.
{\em J. Mech. Phys. Solids} {\bf 41}, 389--412 (1993).

\bibitem{JB93a}
Johnson, M.A., Beatty, M.F.:
The Mullins effect in uniaxial extension and its influence on the
transverse vibration of a rubber string.
{\em Continuum Mech. Thermodyn.} {\bf 5}, 83--115 (1993).

\bibitem{JB93b}
Johnson, M.A., Beatty, M.F.:
A constitutive equation for the Mullins effect in stress controlled
uniaxial extension experiments.
{\em Continuum Mech. Thermodyn.} {\bf 5}, 301--318 (1993).

\bibitem{LAP93}
Lee, B.J., Argon, A.S., Parks, D.M., Ahzi, S., Bartczak, Z.:
Simulation of large strain plastic deformation and texture evolution
in high density polyethylene.
{\em Polymer}, {\bf 34}, 3555--3575 (1993).

\bibitem{MTD93}
Mukhopadhyay, K., Tripathy, D.K., De, S.K.:
Dynamic mechanical properties of silica-filled ethylene vinyl acetate
rubber.
{\em J. Appl. Polym. Sci.} {\bf 48}, 1089--1103 (1993).

\bibitem{TRK93}
Le Tallec, P., Rahier, C., Kaiss, A.:
Three-dimensional incompressible viscoelasticity in large strains:
formulation and numerical approximation.
{\em Comp. Meths. Appl. Mech. Engng.} {\bf 109}, 233--258 (1993).

\bibitem{WRC93}
Witten, T.A., Rubinstein, M., Colby, R.H.:
Reinforcement of rubber by fractal aggregates.
{\em J. Phys. I$\!$I France} {\bf 3}, 367--383 (1993).

\bibitem{BB94}
van den Bogert, P.A.J., de Borst, R.:
On the behavior of rubberlike materials in compression and shear.
{\em Arch. Appl. Mech.} {\bf 64}, 136--146 (1994).

\bibitem{HS95}
Hausler, K., Sayir, M.B.:
Nonlinear viscoelastic response of carbon black reinforced rubber
derived from moderately large deformations in torsion.
{\em J. Mech. Phys. Solids} {\bf 43}, 295--318 (1995).

\bibitem{Ulm95}
Ulmer, J.C.:
Strain dependence of dynamic mechanical properties of carbon black-filled
rubber compounds.
{\em Rubber Chem. Technol.} {\bf 69}, 15--47 (1995).

\bibitem{HS96}
Holzapfel, G., Simo, J.:
A new viscoelastic constitutive model for continuous media
at finite thermomechanical changes.
{\em Int. J. Solids Structures} {\bf 33}, 3019--3034 (1996).

\bibitem{HVH96}
Huber, G., Vilgis, T.A., Heinrich, G.:
Universal properties in the dynamic deformation of filled rubbers.
{\em J. Phys.: Condens. Matter} {\bf 8}, L409--L412 (1996).

\bibitem{Lio96}
Lion, A.:
A constitutive model for carbon black filled rubber:
experimental investigation and mathematical representation.
{\em Continuum Mech. Thermodyn.} {\bf 8}, 153--169 (1996).

\bibitem{Lio97a}
Lion, A.:
A physically based method to represent the thermo-mechanical
behaviour of elastomers.
{\em Acta Mech.} {\bf 123}, 1--25 (1997).

\bibitem{Lio97b}
Lion, A.:
On the large deformation behaviour of reinforced rubber at
different temperatures.
{\em J. Mech. Phys. Solids} {\bf 45}, 1805--1834 (1997).

\bibitem{RW97}
Reese, S., Wriggers, P.:
A material model for rubber-like polymers exhibiting plastic deformation:
computational aspects and comparison with experimental results.
{\em Comp. Meths. Appl. Mech. Engng.} {\bf 148}, 279--298 (1997).

\bibitem{Spa97}
Spathis, G.:
Non-linear constitutive equations for viscoelastic behaviour of
elastomers at large deformations.
{\em Polym. Gels Networks} {\bf 5}, 55--68 (1997).

\bibitem{ZB97}
Zaroulis, J.S., Boyce, M.C.:
Temperature, strain rate, and strain state dependence of the evolution
in mechanical behaviour and structure of poly(ethylene terephthalate)
with finite strain deformation.
{\em Polymer} {\bf 38}, 1303--1315 (1997).

\bibitem{KR98}
Kaliske, M., Rothert, H.:
Constitutive approach to rate-independent properties of filled
elastomers.
{\em Int. J. Solids Structures} {\bf 35}, 2057--2071 (1998).

\bibitem{Lio98}
Lion, A.:
Thixotropic behaviour of rubber under dynamic loading histories:
experiments and theory.
{\em J. Mech. Phys. Solids} {\bf 46}, 895--930 (1998).

\bibitem{RG98}
Reese, S., Govidjee, S.:
Theoretical and numerical aspects in the thermo-viscoelastic material
behaviour of rubber-like polymers.
{\em Mech. Time-Dependent Mater.} {\bf 1}, 357--396 (1998).

\bibitem{SE98a}
Septanika, E.G., Ernst, L.J.:
Application of the network alteration theory for the modeling
the time-dependent constitutive behavior of rubbers.
1. General theory.
{\em Mech. Mater.} {\bf 30}, 253--263 (1998).

\bibitem{SE98b}
Septanika, E.G., Ernst, L.J.:
Application of the network alteration theory for the modeling
the time-dependent constitutive behavior of rubbers.
2. Experimental verification.
{\em Mech. Mater.} {\bf 30}, 255--273 (1998).

\bibitem{LB99}
Llana, P.G., Boyce, M.C.:
Finite strain behavior of poly(ethylene terephthalate) above the
glass transition temperature.
{\em Polymer} {\bf 40}, 6729--6751 (1999).

\bibitem{BSL00}
Boyce, M.C., Socrate, S., Llana, P.G.:
Constitutive model for the finite deformation stress--strain behavior
of poly(ethylene terephthalate) above the glass transition.
{\em Polymer} {\bf 41}, 2183--2201 (2000).

\bibitem{TBP00}
Tzika, P.A., Boyce, M.C., Parks, D.M.:
Micromechanics of deformation in particle-toughened polyamides.
{\em J. Mech. Phys. Solids} {\bf 48}, 1893--1929 (2000).

\bibitem{KB90}
Kramer, H.H., Berger, L.L.:
Fundamental processes of craze growth and fracture.
{\em Adv. Polym. Sci.} {\bf 91--92}, 1--68 (1990).

\bibitem{SG99}
Steenbrink, A.C., van der Giessen, E.:
On cavitation, post-cavitation and yield in amorphous polymer-rubber
blends.
{\em J. Mech. Phys. Solids} {\bf 47}, 843--876 (1999).

\bibitem{DB00}
Dorfmann, A., Burtscher, S.L.:
Aspects of cavitation damage in seismic bearings.
{\em J. Struct. Engng.} {\bf 126}, 573--579 (2000).

\bibitem{HMS00}
Hobeika, S., Men, Y., Strobl, G.:
Temperature and strain rate independence of critical strains in
polyethylene and poly(ethylene-{\it co}-vinyl acetate).
{\em Macromolecules} (in press).

\bibitem{GS97}
Gaucher-Miri, V., Seguela, R.:
Tensile yield of polyethylene and related copolymers: mechanical
and structural evidences of two thermally activated processes.
{\em Macromolecules} {\bf 30}, 1158--1167 (1997).

\bibitem{HHL98}
Hiss, R., Hobeika, S., Lynn, C., Strobl, G.:
Network stretching, slip processes and fragmentation of crystallites
during uniaxial drawing of polyethylene and related copolymers.
A comparative study.
{\em Macromolecules} {\bf 32}, 4390--4403 (1998).

\bibitem{WG93}
Wu, P.D., van der Giessen, E.:
On improved network models for rubbery elasticity and their applications
to orientational hardening in glassy polymers.
{\em J. Mech. Phys. Solids} {\bf 41}, 427--456 (1993).

\bibitem{Tre75}
Treloar, L.R.G.:
The Physics of Rubber Elasticity.
Oxford: Clarendon Press 1975.

\bibitem{Fer80}
Ferry, J.D.:
Viscoelastic Properties of Polymers.
New-York: Wiley 1980.

\bibitem{DE86}
Doi, M., Edwards, S.F.:
The Theory of Polymer Dynamics.
Oxford: Oxford University Press 1986.

\bibitem{AG65}
Adam, G., Gibbs, J.H.:
On the temperature dependence of cooperative relaxation properties
in glass-forming liquids.
{\em J. Chem. Phys.} {\bf 43}, 139--146 (1965).

\bibitem{Dyr95}
Dyre, J.C.:
Energy master equation: a low temperature approach to B\"{a}ssler's
random-walk model.
{\em Phys. Rev. B} {\bf 51}, 12276--12294 (1995).

\bibitem{MB96}
Monthus, C., Bouchaud, J.-P.:
Models of traps and glass phenomenology.
{\em J. Phys. A: Math. Gen.} {\bf 29}, 3847--3869 (1996).

\bibitem{Sol98}
Sollich, P.:
Rheological constitutive equation for a model of soft glassy materials.
{\em Phys. Rev. E} {\bf 58}, 738--759 (1998).

\bibitem{Str90}
Struik, L.C.E.:
Internal Stresses, Dimensional Instabilities and Molecular Orientations
in Plastics.
Chichester: Wiley 1990.

\bibitem{GT46}
Green, M.S., Tobolsky, A.V.:
A new approach to the theory of relaxing polymeric media.
{\em J. Chem. Phys.} {\bf 14}, 80--92 (1946).

\bibitem{Yam56}
Yamamoto, M.:
The visco-elastic properties of network structure.
1. General formalism.
{\em J. Phys. Soc. Japan} {\bf 11}, 413--421 (1956).

\bibitem{Lod68}
Lodge, A.S.:
Constitutive equations from molecular network theories
for polymer solutions.
{\em Rheol. Acta} {\bf 7}, 379--392 (1968).

\bibitem{TE92}
Tanaka, F., Edwards, S.F.:
Viscoelastic properties of physically cross-linked networks.
Transient network theory.
{\em Macromolecules} {\bf 25}, 1516--1523 (1992).

\bibitem{Wan92}
Wang, S.-Q.:
Transient network theory for shear-thickening fluids and physically
cross-linked systems.
{\em Macromolecules} {\bf 25}, 7003--7010 (1992).

\bibitem{SJB00}
Serero, Y., Jacobsen, V., Berret, J.-F., May, R.:
Evidence of nonlinear chain stretching in the rheology of
transient networks.
{\em Macromolecules} {\bf 33}, 1841--1847 (2000).

\bibitem{PTT77}
Phan-Thien, N., Tanner, R.I.:
A new constitutive equation derived from network theory.
{\em J. Non-Newtonian Fluid Mech.} {\bf 2}, 353--365 (1977).

\bibitem{FL81}
Fuller, G.G., Leal, L.G.:
Network models of concentrated polymer solutions derived from the
Yamamoto network theory.
{\em J. Polym. Sci.: Polym. Phys. Ed.} {\bf 19}, 531--555 (1981).

\bibitem{PB88}
Petruccione, F., Biller, P.:
Rheological properties of network models with con\-fi\-guration-dependent
creation and loss rates.
{\em Rheol. Acta} {\bf 27}, 557--560 (1988).

\bibitem{Pal97}
Palierne, J.-F.:
Sticky dumbbells: from Hookean dumbbells to transient network.
{\em Rheol. Acta} {\bf 36}, 534--543 (1997).

\bibitem{BS99}
Barsky, S., Slater, G.W.:
A nonequilibrium molecular dynamic simulation of the time-dependent
orientational coupling between long and short chains in a bimodal
polymer melt upon uniaxial stretching.
{\em Macromolecules} {\bf 32}, 6348--6358 (1999).

\bibitem{AO95}
Ahn, K.H., Osaki, K.:
Mechanism of thear thickening investigated by a network model.
{\em J. Non-Newtonian Fluid Mech.} {\bf 56}, 267--288 (1995).

\bibitem{Mor98a}
Morse, D.C.:
Viscoelasticity of concentrated isotropic solutions of semiflexible
polymers.
1. Model and stress tensor.
{\em Macromolecules} {\bf 31}, 7030--7043 (1998).

\bibitem{Mor98b}
Morse, D.C.:
Viscoelasticity of concentrated isotropic solutions of semiflexible
polymers.
2. Linear response.
{\em Macromolecules} {\bf 31}, 7044--7067 (1998).

\bibitem{Dro96}
Drozdov, A.D.:
Finite Elasticity and Viscoelasticity.
Singapore: World Scientific 1996.

\bibitem{CG67}
Coleman, B.D., Gurtin, M.E.:
Thermodynamics with internal state variables.
{\em J. Chem. Phys.} {\bf 47}, 597--613 (1967).

\bibitem{AD90}
Altenberger, A.R., Dahler, J.S.:
Statistical mechanics of rubber elasticity.
{\em J. Chem. Phys.} {\bf 92}, 3100--3111 (1990).

\bibitem{Eve95}
Everaers, R.:
Elasticity of $c^{\ast}$-gels.
{\em J. Phys. I$\!$I France} {\bf 5}, 1491--1500 (1995).

\bibitem{Moo40}
Mooney, M.:
A theory of large elastic deformation.
{\em J. Appl. Phys.} {\bf 11}, 582--592 (1940).

\bibitem{RS51}
Rivlin, R.S., Saunders, D.W.:
Large elastic deformations of isotropic materials.
7. Experiments on the deformation of rubber.
{\em Phys. Trans. Roy. Soc. London} {\bf 243}, 251--288 (1951).

\bibitem{ZMF86}
Zang, Y.H., Muller, R., Froelich, D.:
Interpretation of the rheological behaviour in elongation
of uncrosslinked polystyrene melts in terms of the Mooney--Rivlin
equation.
{\em Polymer} {\bf 27}, 61--65 (1986).

\bibitem{NR94}
Ngai, K.L., Roland, C.M.:
Junction dynamics and the elasticity of networks.
{\em Macromolecules} {\bf 27}, 2454--2459 (1994).

\bibitem{GCZ96}
Goldbart, P.M., Castillo, H.E., Zippelius, A.:
Randomly crosslinked macromolecular systems: vulcanization transition
to and properties of the amorphous solid state.
{\em Adv. Phys.} {\bf 45}, 393--468 (1996).

\bibitem{KE75}
Krausz, A.S., Eyring, H.:
Deformation Kinetics.
New York: Wiley, 1975.
\end{thebibliography}
\end{document}